\documentstyle[12pt]{article}
\def\appendix#1{
\addtocounter{section}{1}
\setcounter{equation}{0}
\renewcommand{\thesection}{\Alph{section}}
\section*{Appendix \thesection\protect\indent #1}
\addcontentsline{toc}{section}{Appendix \thesection\ \ \ #1}
}

\def\be{\begin{equation}}
\def\la{\label}
\def\ee{\end{equation}}
\def\bea{\begin{eqnarray}}
\def\eea{\end{eqnarray}}
\def\eps{\varepsilon}
\def\a{\alpha}
\def\b{\beta}

\def\n{\nabla}

\def\d{\delta}

\def\p{\partial}

\begin{document}
\title{On the origin of supergravity boundary terms in the AdS/CFT
correspondence} \author{G.E.Arutyunov\thanks{arut@genesis.mi.ras.ru}
\mbox{} and S.A.Frolov
\thanks{
Address after September 1, 1998: The University of Alabama, Department
of Physics and Astronomy, Box 870324, Tuscaloosa, Alabama 35487-0324.
}
\mbox{} \\
\vspace{0.4cm}
Steklov Mathematical Institute,
\vspace{-0.5cm} \mbox{} \\
Gubkin str.8, GSP-1, 117966, Moscow, Russia;
\vspace{0.5cm} \mbox{}
}
\date {}
\maketitle
\begin{abstract}
The standard formulation of the AdS/CFT correspondence
is incomplete since it requires adding to a supergravity
action some a priori unknown boundary terms.
We suggest a modification of the correspondence principle
based on the Hamiltonian formulation of the supergravity action,
which does not require any boundary terms.
Then all the boundary terms of the standard formulation naturally
appear by passing from the Hamiltonian version to the Lagrangian one.
As examples the graviton part of the supergravity action on the product
of $AdS_{d+1}$ with a compact Einstein manifold $\cal E$ and fermions
on $AdS_{d+1}$ are considered. We also discuss conformal transformations
of gravity fields on the boundary of $AdS$ and show that they are
induced by the isometries of $AdS$.
\end{abstract}

\section{Introduction}
According to a recent conjecture by Maldacena \cite{M} the large $N$
limit of certain conformal field theories in $d$ dimensions can be governed
by supergravity on the product of $d+1$-dimensional anti de Sitter
space $AdS_{d+1}$ with a compact manifold ${\cal E}$. This conjecture
was further elaborated by Gubser, Klebanov and Polyakov \cite{GKP}
and Witten \cite{W}, who proposed to identify the generating
functional of the connected Green functions in a $d$-dimensional
conformal field theory with the minimum of the supergravity action
subject to certain conditions imposed on supergravity fields
at the boundary of $AdS_{d+1}\times {\cal E}$. In \cite{AV}-\cite{SSRS}
the proposal was used to compute some two- and three-point correlation
functions in ${\cal N}=4$ $SU(N)$ Yang-Mills theory, which exhibit
the expected conformal behavior. However, to make the AdS/CFT
correspondence complete one has to add to the supergravity action some
boundary terms.

The necessity of the boundary terms can be seen as follows.
Firstly, to calculate the Green functions in field theory
one has to introduce the UV regularization, for example, the lattice
regularization with the lattice step $a$. Since any regularization
violates the conformal invariance, one has to add some counterterms
to the field theory action, to restore (if possible) the conformal
invariance  in the limit $a\to 0$.

Secondly, recall that the geometry of the $AdS_{d+1}$ space is described
by the metric
\bea
ds^2=\frac{1}{x_0^2}(\eta_{ij}dx^idx^j+dx_0^2),
\nonumber
\eea
where $\eta_{ij}$ is the $d$-dimensional Minkowski metric.
The boundary of $AdS_{d+1}$ space is at $x_0=0$ and can be identified
with the Minkowski space. The boundary is located at infinite distance
from any point of $AdS_{d+1}$. This leads to infrared divergency of
the gravity action. To make the action finite
\footnote{Strictly speaking, the
regularized action is still infinite because of the infinite volume
of the boundary. This infinity is cured up by adding proper boundary
terms.}
one should cut $AdS_{d+1}$ space off at $x_0=\eps$ and
consider the part of $AdS_{d+1}$ with $x_0\ge \eps$. The minimum of
the gravity action is calculated by requiring the gravity fields to
vanish at $x_0=\infty$ and to take arbitrary values at $x_0=\eps$.

The IR regularization leads to explicit violation of the conformal
invariance of the gravity action
and to restore it in the limit $\eps\to 0$ one has to add
some boundary terms.
In fact, the gravity action considered as a functional of the
boundary fields is never conformal invariant \cite{Aref}.
The conformal invariance of the $AdS_{d+1}$ gravity should be understood
as the validity of the conformal Ward identities for correlation
functions\footnote{
Recently an attempt \cite{KvP} to present a general proof of
conformal invariance of the CFT Green functions computed from
supergravity on $AdS$ background was made. However, the arguments
of \cite{KvP} do not take into account IR divergencies and rely on
inproper transformation laws of boundary fields.}.
It is clear from this discussion that the UV
regularization and the counterterms of the field theory correspond to
the IR regularization and the boundary terms of the gravity theory,
the UV cutoff $a$ being identified with the IR cutoff $\eps$.

Then the standard formulation of the
AdS/CFT correspondence  \cite{GKP,W} asserts that a connected
Green function in the $d$-dimensional conformal field theory is equal
to
\be
\lim_{a\to 0} \langle {\cal O}_1(x_1)\ldots{\cal
O}_n(x_n)\rangle = \lim_{\eps\to 0}\frac{\delta}{\delta\phi_1
(x_1)}\cdots\frac{\delta}{\delta\phi_n (x_n)}
\mbox{min}\,S_{gr}(\phi_1 ,\ldots ,\phi_n )|_{\phi_i
(x,x_0=\eps)=\phi_i(x)}
\la{AdSCFT}
\ee
where ${\cal O}_i(x_i)$ are some gauge-invariant operators in the
field theory and $\phi_i(x_i)$ are the gravity fields corresponding
to ${\cal O}_i(x_i)$. The gravity action
$S_{gr}(\phi_1 ,\ldots ,\phi_n )$ is the sum of the bulk action
and the boundary terms. Note that the boundary terms that depend
locally on the boundary fields $\phi_i(x)$ (i.e. do not depend on $\p_0\phi_i(x)$)
do not contribute to the Green functions.
From the viewpoint of the boundary CFT
these local boundary terms reflect the
usual ambiguity in the definition of the $T$-product.
Since the conformal
invariant correlation functions are not distributions, one can not
use the Fourier image of the Green functions in eq.(\ref{AdSCFT}).
Loosely speaking, the generating functional of the Green functions
coincides with the partition function of the supergravity (string) theory.

It is worth noting that since the boundary terms are in general unknown,
the usual formulation of the AdS/CFT correspondence is incomplete.
Nevertheless, it seems possible to formulate the AdS/CFT correspondence
in a way that does not depend on any unknown boundary terms.
To this end remind that in quantum mechanics the minimum of an effective
action with given initial $q_{in}$ and final $q_f$ configurations
is proportional to the logarithm of the transition amplitude from
the initial state $|q_{in}>$ to the final state $|q_f>$. Note that the
effective action one gets from the transition amplitude is always
of the form $\int d t (p\dot{q}-H(p,q))$ without any boundary terms
since the initial and final states are taken in the coordinate
representation. Coming back to the AdS/CFT correspondence
we see that the coordinate $x_0$ plays a distinguished role.
If we introduce a new coordinate $t=\log x_0$ and interpret
$t$ as the time coordinate in $AdS$, then we can postulate
that the generating functional of the Green functions coincides
with the string transition amplitude from the initial state $|\phi>$
to the "vacuum" state $|0>$:
\bea
\nonumber
Z(\phi)=\lim_{{t'\to -\infty \atop t''\to +\infty}}
<0|\mbox{e}^{-i\int_{t'}^{t''} H_{str}(t)dt}|\phi>,
\eea
where $H_{str}$ is the string theory Hamiltonian governing the
dynamics in the $t$ direction.

In the classical supergravity limit one recovers the above-mentioned
version of the AdS/CFT correspondence with the supergravity boundary terms
fixed by the Hamiltonian formulation of the supergravity action. It means
that all the boundary terms naturally appear by passing from the Hamiltonian
version to the Lagrangian one.
In particular, a local boundary term appears in the bulk as a total derivative term.
From the Hamiltonian point of view such a term can be removed by
performing a canonical transformation affecting only the momenta.
However, a choice of momenta is partially fixed by requiring
the minimum of the Hamiltonian to be achieved at $p=0=\phi$. Such a choice
of momenta would ensure the vanishing of the one-point correlation functions.

Since supergravity is a constrained system,
to obtain its Hamiltonian formulation one has to impose some gauge conditions
and to eliminate the second-class constraints. The most convenient
gauges seem to be a generalization of the temporal gauge $A_0=0$ in the
Yang-Mills theory. In these gauges the gravity action has a residual
symmetry with gauge parameters that do not depend on $t$. The residual
symmetry leads to conservation laws for currents of the boundary
field theory. Moreover, on the boundary of $AdS$ the residual gauge
group gets enhanced and includes, in particular, isometries
of the background space. In the paper we show that conformal transformations
of the supergravity boundary fields are induced by the isometries.

In this paper we demonstrate how the Hamiltonian version of the
AdS/CFT correspondence permits to determine the supergravity boundary
terms that have to be added to the standard supergravity action.

The plan of the paper is as follows. In the second Section we consider
a gravity model on the $AdS\times {\cal E}$ background, where ${\cal E}$
is an arbitrary Einstein manifold. We show that the Hamiltonian formulation
correctly predicts the standard boundary term \cite{GH} as well as
the term proportional to the volume of the boundary suggested in \cite{T}
for the case of the gauged supergravity on $AdS_{d+1}$
describing the dynamics of zero modes in $AdS_{d+1}\times {\cal E}$
compactification. There is also an additional term which, however, is a
local functional of boundary fields and, thereby, does not change
the value of the Green functions.
The boundary terms discussed in the second Section can be equivalently
found relying on the conformal invariance of the corresponding
boundary field theory.
Namely, the conformal invariance requires any one-point correlation function
to vanish, and, therefore, the minimum of the supergravity action
extended over the $AdS_{d+1}\times {\cal E}$ background should not have
linear dependence on the boundary values of the supergravity fields.
In this Section we also consider the conformal transformations
of the boundary graviton induced by the isometries of $AdS$.

In the third Section we discuss fermions on the $AdS$ background and
demonstrate that the fermion boundary term suggested in \cite{HS}
again follows from the Hamiltonian version of the action. We also
study conformal transformations of the boundary fermion fields.

\section{Metric dependent boundary terms}
Clearly, the boundary terms depending only on the metric
originate from the Einstein-Hilbert term
\bea
\label{bulk}
I=\int_{M} d^Dx \sqrt{-g}R,
\eea
where we choose the gravitational constant equal to one.
We consider the theory on a manifold with a boundary $\p M$.
By using the reparametrization invariance of action (\ref{bulk})
we can describe $\p M$ at least locally as a
hypersurface $t=\log\eps=const$ and choose
in the vicinity of $\p M$ a gauge $g_{tt}=1$ and $g_{t\mu}=0$.

Remind that to study the Hamiltonian formulation it is useful to represent
$\sqrt{-g}R$ in the following form
\bea
\la{R}
\sqrt{-g}R=\sqrt{-g}g^{ab}(\Gamma_{ad}^c\Gamma_{cb}^d-
\Gamma_{ab}^c\Gamma_{cd}^d)
+\p_a(\sqrt{-g}g^{bc}\Gamma^a_{bc}-\sqrt{-g}g^{ab}\Gamma^c_{bc}).
\eea
The last term is a total derivative term and should be omitted in
the Hamiltonian formulation.

Considering a manifold with a boundary one usually adds to
(\ref{bulk}) the standard boundary term
that removes all terms linear in second derivatives \cite{GH}:
\bea
\label{sbt}
I^{(1)}=2\int_{\p M} d^{D-1}x \sqrt{-\bar{g}}K.
\eea
Here $K$ is a trace of the second fundamental form on the boundary and
$\bar{g}$ is the determinant of the induced metric.
Remind that introducing the normal $n^a$ to the boundary, $K$ can
be written as $K=\n_a n^a$. In our gauge
the induced metric coincides with $g_{\mu\nu}$, $g=\bar{g}$
and $n^a=(-1,0,\ldots ,0)$. Therefore,
\bea
\la{K}
I^{(1)}=2\int_{\p M} d^{D-1}x
\sqrt{-\bar{g}}\frac{1}{\sqrt{-g}}\p_a(\sqrt{-g}n^a)=
-2\int_{\p M} d^{D-1}x \p_t(\sqrt{-\bar{g}}).
\eea
Now one can easily see that integrating
the total derivative term in (\ref{R}) one obtains
$-I^{(1)}$.
Thus, adding to (\ref{bulk}) boundary
term $I^{(1)}$ we get the action that should be used to obtain
the Hamiltonian formulation. It is worth noting that $I^{(1)}$
is not cancelled by the total derivative term in (\ref{bulk})
in an arbitrary gauge. However, the difference is always a local functional
of boundary fields and, therefore, is irrelevant for computing
the Green functions.

Thus, the action $S=I+I^{(1)}$ acquires a form
\bea
\la{bul}
S=\int_{M} d^Dx\sqrt{-g}\left(\frac{1}{4}\p_tg^{\mu\nu}\p_tg_{\mu\nu}
+\frac{1}{4}g^{\mu\nu}\p_tg_{\mu\nu} g^{\rho\lambda} \p_tg_{\rho\lambda}
+g^{\mu\nu}(\Gamma_{\mu\rho}^\lambda\Gamma_{\lambda\nu}^\rho-
\Gamma_{\mu\nu}^\lambda\Gamma_{\lambda\rho}^\rho)\right).
\eea
Suppose that a vacuum solution of the Einstein equations defines the
following metric
\bea
ds^2=dt^2+G_{\mu\nu}dx^{\mu}dx^{\nu}=
dt^2+e^{-2t}\eta_{ij}dx^idx^j+G_{\a\b}dx^{\a}dx^{\b}
\la{met}
\eea
on the manifold $M=AdS_{d+1}\times {\cal E}_p$, where ${\cal E}_p$ is
a compact Einstein manifold with the metric $G_{\a\b}$:
$R_{\a\b}=\nu G_{\a\b}$. Here $i,j=1,\ldots, d$; $\a,\b=1,\ldots, p$,
$D=d+1+p$ and $\eta_{ij}$ is the metric of the $d$-dimensional Minkowski space.
The boundary of $M$ is a hypersurface $t=\log\eps=const$, which can be
considered as the product of the Minkowski space with ${\cal E}_p$.

It is convenient to decompose the metric as follows
$g_{\mu\nu}=G_{\mu\rho}H_{\nu}^{\rho}$. Then, (\ref{bul}) reads as
\bea
\la{buh}
S&=&\int_{M} d^Dx\sqrt{-G}\sqrt{H}
\left(-d(d-1)+\frac{1}{4}\p_tH_{\mu}^{\nu}\p_t(H^{-1})_{\nu}^{\mu}
+\frac{1}{4}(\p_tH_{\mu}^{\nu}(H^{-1})_{\nu}^{\mu})^2 \right.\\
\nonumber
&-&\left.
\p_tH_{\mu}^{\a}(H^{-1})_{a}^{\mu}
+g^{\mu\nu}(\Gamma_{\mu\rho}^\lambda\Gamma_{\lambda\nu}^\rho-
\Gamma_{\mu\nu}^\lambda\Gamma_{\lambda\rho}^\rho)\right)
+2(1-d)\int_{M}d^Dx\p_t\sqrt{-g}.
\eea
The last term is a total derivative term and
should be omitted in passing to the Hamiltonian version.
From the Lagrangian point of view this total derivative term can be
cancelled by adding to (\ref{buh}) the boundary term proportional
to the volume of the boundary:
\bea
\la{area}
I^{(2)}=2(1-d)\int_{\p M}\sqrt{-\bar{g}}.
\eea

As was mentioned in the Introduction one should require the minimum
of the Hamiltonian to be achieved at zero value of coordinates and momenta.
The only term that could shift the value of momenta is
$-\sqrt{-g}\p_tH_{\mu}^{\a}(H^{-1})_{a}^{\mu}$.
Indeed, expanding this term near the background one gets
$$-\p_t(\sqrt{-G} h_{\a}^{\a})-d\sqrt{-G} h_{\a}^{\a}+O(h^2),$$
where $H_{\mu}^{\nu}=\d_{\mu}^{\nu}+h_{\mu}^{\nu}$. The first term in this
expansion is a total derivative term linear in $h$ and, therefore,
shifts the value of momenta. It turns out that from the Lagrangian
point of view there exists a lot of covariant boundary terms that
can compensate this shift. For example, one can add to action (\ref{buh})
the following term
\bea
\int_{\p M}\sqrt{-\bar{g}}f(\bar{R}),
\eea
where $\bar{R}$ is the curvature of the metric induced on the boundary.
The function $f(x)$ should satisfy two conditions: $f(\nu p)=0$ and
$f'(\nu p)=1/\nu$. A possible choice is $f(x)=x-\frac{1}{\nu p}x^2$.
Thus, the action that admits a straightforward Hamiltonization
and should be used in computation of the Green functions has the form
\bea
\la{bound}
{\bf S}=\int_{M} d^Dx \sqrt{-g}R+
2\p_n\int_{\p M}d^{D-1}x \sqrt{\bar{g}} +
2(1-d)\int_{\p M}d^{D-1}x \sqrt{\bar{g}}+
\int_{\p M}d^{D-1}x \sqrt{\bar{g}}f(\bar{R}).
\eea

To compare the boundary terms (\ref{bound}) with the ones introduced
for the pure $AdS_{d+1}$ gravity \cite{T} one obviously has to
perform the dimensional reduction of ${\bf S}$ to
$D=d+1$ dimensions. Leaving only the graviton
modes one gets
\bea
\nonumber
{\bf S}=\int_{M} d^{d+1}x \sqrt{-g}(R+d(d-1))
+2\p_n\int_{\p M}d^dx \sqrt{\bar{g}} +
2(1-d)\int_{\p M}d^dx \sqrt{\bar{g}}+
\int_{\p M}d^dx \sqrt{\bar{g}}f(\bar{R}), \\
\la{b}
\eea
where an overall multiplier being the  volume of the sphere is omitted.
Here $d(d-1)$ is the cosmological constant needed to ensure the
consistency of the reduction.
The first two boundary terms are exactly the same as in \cite{T},
where they were added to cancel boundary terms that are linear
in metric perturbation of the action of $D=d+1$ dimensional gravity near
$AdS_{d+1}$ background. Since $f(0)=0$ after the reduction  and
$\bar{R}=0$ on the background  the third boundary term
does not contribute to the first variation of (\ref{b}).
The boundary area counter-term is needed just for
cancelling the linear terms ensuring thereby the vanishing of one-point
Green functions. Only the second fundamental form is important for
computing the higher-point Green functions. The other boundary
terms are local and as was discussed in the Introduction
do not contribute.

Now we discuss the conformal transformations of the graviton on
the boundary induced by isometries of $AdS$. In the sequel
it is convenient to deal with the coordinate $x_0=\mbox{e}^t$ and
to decompose $g_{ab}=G_{ab}+h_{ab}$.
Denote by $\xi^a$ a Killing vector
of the background metric. Assuming $G$ to be stationary
under diffeomorphism generated by $\xi$ one finds that transformation
\bea
\d h_{ab}=\xi^c\p_c h_{ab}+h_{ac}\p_b\xi^c+h_{bc}\p_a\xi^c
\la{sym}
\eea
is a rigid symmetry of equations of motion in every order in
metric perturbation $h_{ab}$.
Note that the Killing vectors of the $AdS_{d+1}$ background can
be written as
\bea
\la{Kil}
&&\xi^0=x_0(A_kx^k+D),\\
\nonumber
&&\xi^i=-\frac{x_0^2-\eps^2}{2}A^i+\left(
-\frac{1}{2}(A^i x^2 -2 x^i A_k x^k)+Dx^i+\Lambda_j^i x^j+P^i
\right),
\eea
where $A^i,D,\Lambda_j^i,P^i$ generate on the boundary special
conformal transformations, dilatations, Lorentz transformations and shifts
respectively.
Transformations (\ref{sym}) do not respect the gauge
conditions $h_{0i}=0$: $\d h_{0i}=h_{ij}\p_0\xi^j$.
However, the gauge can be restored by combining (\ref{Kil})
with the transformation
$\d h_{ab}=\n_a\chi_b+\n_b\chi_a$ with $\chi^0=0$ and
$\chi^i=\int_{\eps}^{x_0}zg^{kj}h_{ij}(z,\vec{x})A_k$. Now it is easy to find
\bea
\la{tra}
\d h_i^j= \xi^k\p_k h_i^j+h_i^k\p^j\xi_k+h_k^j\p_i\xi^k+\xi^0\p_0h_i^j
-\frac{2}{x_0}\xi^0h_i^j +\p_i\chi^j+\p^i\chi_j,
\eea
where $h_i^j=G^{jk}h_{ik}$.

Note that on the boundary $\chi^i$ vanish. Clearly, to obtain
the induced transformations on the boundary one should know the
behavior of $\p_0h_i^j$ in the limit $\eps\to 0$.
To this end we consider the linearized equations of motion following
from (\ref{b}):
\bea
\n_k\n^k h_{ab}+\n_a\n_b h -\n_a\n_c h_b^c -\n_b\n_c h_a^c+
2(h_{ab}-g_{ab}h)=0.
\la{eqm}
\eea
Eq.(\ref{eqm}) has the standard symmetry
$\d h_{ab}=\n_a\xi_b+\n_b\xi_a$
that allows one to impose the gauge
conditions $h_{0a}=0$.
Below, to simplify the notation, we adopt the convention
that the indices $i,j$ are raised and lowered by using the Minkowski
metric $\eta_{ij}$, in particular, $\Box=\eta^{ij}\p_i\p_j$.
In this gauge covariant equation (\ref{eqm})
written for $h_i^j$ acquires the form
\bea
\la{eqm1}
\p_0^2 h_i^j+\Box h_i^j+\frac{1-d}{x_0}\p_0h_i^j
-\frac{1}{x_0}\d_i^j\p_0 h
+\left(\p_i\p^j h
-\p_i\p^k h_k^j-\p^j\p_k h_k^i\right)=0.
\eea
In addition one has two constraints:
\bea
\la{con1}
&&\p_0\left(\p_i h - \p_k h_i^k\right) =0,\\
\la{con2}
&&\Box h+\frac{(1-d)}{x_0}\p_0 h - \p_i\p^j h^i_j=0,
\eea
which follow from the equation for $h_{0a}$.
Introduce the transversal part $h^{\perp~j}_i$ of $h_i^j$:
\bea
h^{\perp~j}_i\equiv h_i^j-\frac{1}{\Box}\p_i\p^kh_k^j-\frac{1}{\Box}
\p^j\p_k h^k_i +\frac{\p_i\p^j}{\Box^2}\p_k\p^m h_m^k,
\la{perp}
\eea
The trace $h^{\perp}$ of the transversal part satisfies
$\Box(h-h^{\perp})=\p_i\p^jh^i_j$.

Obviously, by using (\ref{con1}) one gets
$$
\p_0 h=\p_0 h^{\perp}+\frac{1}{\Box}\p^i\p_0\p_j h_i^j=
\p_0 h^{\perp}+\p_0 h.
$$
Thus,
\bea
\la{bas}
\p_0 h^{\perp}=0,~~~~\p_0 h=\frac{x_0}{d-1}\Box h^{\perp},
\eea
where the last equation follows immediately from eq.(\ref{con2}).

Introduce a traceless transversal part $\bar{h}_i^j$ of $h_i^j$:
\bea
\la{ttp}
\bar{h}_i^j=h^{\perp~j}_i+
\frac{1}{d-1}\left(\frac{\p_i\p^j}{\Box}-\d_i^j\right)h^{\perp}.
\eea
Then with the account of (\ref{bas}) eq.(\ref{eqm1}) can be written
as the following equation on $\bar{h}_i^j$:
\bea
\la{scal}
\p_0^2
\bar{h}^j_i+\Box \bar{h}^j_i +\frac{1-d}{x_0}\p_0 \bar{h}^j_i=0,
\eea
which coincides with the equation for a free massless
scalar field on the $AdS$ background.
The Fourier mode solution $\bar{h}^j_i(x_0,\vec{k})$
of eq.(\ref{scal}) obeying the boundary condition
$\bar{h}^j_i(\eps,\vec{k})=h^j_i(\vec{k})$, where $h^j_i(\vec{k})$ is
an arbitrary transversal tensor, reads as
\bea
\la{sol}
\bar{h}^j_i(x_0,\vec{k})=\left(\frac{x_0}{\eps}\right)^{d/2}
\frac{K_{d/2}(x_0k)}{K_{d/2}(\eps k)}h^j_i(\vec{k}),
\eea
where $K_{d/2}$ is the Mackdonald function and $k=|\vec{k}|$.
It is clear from (\ref{sol}) that $\p_0\bar{h}^j_i(\eps,\vec{k})=O(\eps)$.
By using eqs.(\ref{con1}) and (\ref{con2}) one can see
that $\p_0 \p_jh^j_i(\eps,\vec{k})=O(\eps)$ and, therefore,
$\p_0 h^j_i(\eps,\vec{k})=O(\eps)$.

Thus, coming back to eq.(\ref{tra}),
in the limit $\eps\to 0$ we are left with the following
transformation law induced on the boundary
\bea \la{tral}
\d h_i^j(\vec{x})=\xi^k\p_k h_i^j
+h_i^k\p^j\xi_k+h_k^j\p_i\xi^k-2(A_kx^k+D)h_i^j,
\eea
where $\xi^i=\xi^i(\eps,\vec{x})$. This transformation law shows,
in particular, that under dilatations the boundary graviton
transforms with scaling dimension zero as required by the AdS/CFT
correspondence since it couples with the stress-energy tensor whose
scaling dimension is $d$.

As a byproduct of our consideration we can easily derive the two-point
Green function for the stress-energy tensor in the boundary CFT.
Quadratic part of action (\ref{b}) computed on the solution of
equations of motion reduces to the following form
\bea
\la{com1}
{\bf S}^{(II)}=\frac{1}{4}\int d^{d}x \eps^{1-d}
\left(h_i^j\p_0 h_j^i-h\p_0h\right),
\eea
where we have omitted all irrelevant local terms.
Using eqs.(\ref{bas}) it is easy to show that
$$
h_i^j\p_0 h_j^i-h\p_0 h=\bar{h}_i^j\p_0 \bar{h}_j^i-\frac{x_0}{d-1}h^{\perp}\Box h^{\perp},
$$
and, therefore,
\bea
\la{com2}
{\bf S}^{(II)}=\frac{1}{4}\int d^{d}x \eps^{1-d} \bar{h}_i^j\p_0 \bar{h}_j^i,
\eea
where again the local term was omitted.
Since $\bar{h}_i^j$ satisfies the massless scalar field equation
on the $AdS$ background eq.(\ref{com2}) leads to the expected
two-point Green function of the stress-energy tensor.
Note that our computation of the Green function essentially differs
from the one in \cite{T}. In particular, we do not impose the
requirement of vanishing the trace of the graviton.
In \cite{T} $h$ was put equal to zero by arguing that otherwise
it blows up at infinity. However,
eq.(\ref{bas}) is the first order equation and, therefore,
one can not require any condition at infinity. Moreover,
the AdS/CFT correspondence implies the decoupling of $h$ on
the boundary since, otherwise, the trace of the stress-energy tensor
would not be equal to zero. The fact of decoupling of $h$ should
be checked explicitly. In our computation scheme the
contribution of $h$ results only in a local boundary term. In principle,
this contribution can be cancelled by choosing in (\ref{b})
a proper function $f$.

Another representative example demonstrating the relation between
the transformation law in the bulk and the conformal properties of the
boundary data is provided by a massive scalar $\phi$ on $AdS_{d+1}$
background. The equation of motion for $\phi$
\bea
\la{scalm}
\p_0^2 \phi+\frac{1-d}{x_0}\p_0 \phi +(\Box-\frac{m^2}{x_0^2})\phi=0
\eea
has a Fourier mode solution
\bea
\la{sksol}
\phi(x_0,\vec{k})=\left(\frac{x_0}{\eps}\right)^{\frac{d}{2}}
\frac{K_{\nu}(x_0k)}{K_{\nu}(\eps k)}\phi_0(\vec{k}),~~~~
\nu^2=m^2+\frac{d^2}{4}
\eea
that equals to $\phi_0(\vec{k})$ on the boundary. Under
the $AdS$ isometries (\ref{Kil}) the scalar $\phi$ transforms as
\bea
\la{sctr}
\d\phi(x_0,\vec{x})=\xi^a\p_a\phi=\xi^i\p_i\phi+\xi^0\p_0\phi
\eea
and, therefore, the transformation properties of the boundary value
$\phi_0$ are determined by $\lim_{x_0\to \eps} \xi^0\p_0\phi$.
Computing the derivative of $\phi$ in the bulk direction one finds
$\p_0\phi(\eps,\vec{x})=(d/2-\nu)\eps^{-1}\phi_0(\vec{x})+O(1)$ and,
therefore,
\bea
\la{scder}
\d\phi_0(\vec{x})=\xi^i(\vec{x})\p_i\phi_0(\vec{x})+
\left(\frac{d}{2}-\nu\right)(A_kx^k+D)\phi_0(\vec{x}),
\eea
where $\xi^i(\vec{x})=\xi^i(\eps,\vec{x})$
Thus, at the boundary $\phi$ couples with the operator of conformal dimension
$\Delta=d/2+\nu$ as it should be.

These examples show that the conformal properties of boundary data
are not completely determined by the transformation law of the corresponding
bulk fields but depend essentially on the explicit form of the
interaction in the bulk. One could imagine such an exotic interaction
that would lead to a nonlinear or nonlocal realization of the conformal
group on the boundary.

\section{Fermions on the $AdS$ background}
We start with the action for the free massive Dirac fermion
on the $AdS$ background
\bea
\la{sp}
S_{\psi}(\alpha)=\int_M d^{d+1}x\sqrt{-G
}\left(
\frac{\alpha}{2}\bar{\psi}\Gamma^{\mu}D_{\mu}\psi-
\frac{\beta}{2}D_{\mu}\bar{\psi}\Gamma^{\mu}\psi
-m\bar{\psi}\psi\right),
\eea
where $\alpha+\beta=2$. Note that actions with different $\alpha$ and
$\beta$ differ only by a total derivative.
In the sequel we denote by $\mu,\nu,...$ and by $a,b,..$ the world and
the tangent indices respectively, $\Gamma^{\mu}=e^{\mu}_a \gamma^a$
and $\gamma_a$ are the Dirac matrices of $d+1$ dimensional Minkowski
space: $\gamma_a\gamma_b+\gamma_b\gamma_a=2\eta_{ab}$.
Here the conjugated fermion is defined as $\bar{\psi}=\psi^*\gamma^1$ since
$x^1$ is the time direction of the boundary field theory.

Introduce the left and right chiral components of $\psi$:
$\psi_{\pm}=\frac{1}{2}(1\pm \gamma_0)\psi$ and 
define $\bar{\psi}_{\pm}=({\psi}_{\pm})^*\gamma^1$.
Note that $\bar{\psi}_{\pm}\gamma^0=\mp\bar{\psi}_{\pm}$.

Action (\ref{sp})
is already written in the first-order formalism. Therefore,
either $\psi_-$ or $\psi_+$ plays the role of the coordinate.
In order to understand what component of $\psi$ should be regarded
as the coordinate we discuss the transformation properties of $\psi$
at the boundary under isometries of the $AdS$ background.

Recall that under the coordinate transformation generated
by $\xi^{\mu}$ the veilbein $e_{\mu}^a$ transforms as follows
\bea
\d e_{\mu}^a=\xi^{\rho}\n_{\rho}e_{\mu}^a+\n_{\mu}\xi^{\rho}e_{\rho}^a,
\la{ctetr}
\eea
where $\n_{\rho}$ is a covariant derivative with respect to the background
metric $G_{\mu\nu}$. If $\xi^{\mu}$ is a Killing vector of $G_{\mu\nu}$,
then $\n_{\mu}\xi_{\lambda}=\n_{[\mu}\xi_{\lambda]}$.
In addition to (\ref{ctetr}) one also has the local Lorentz symmetry
$\d e_{\mu}^a=\Lambda^a_b(x)e_{\mu}^b$. Thus, the general symmetry
transformation is
\bea
\d e_{\mu}^a=\xi^{\rho}\n_{\rho}e_{\mu}^a+\n_{\mu}\xi^{\rho}e_{\rho}^a+
\Lambda^a_be_{\mu}^b.
\la{tetg}
\eea
However, in a fixed background we are left with the transformations
that preserve $e_{\mu}^a$: $\d e_{\mu}^a=0$. This condition can be attained
by choosing the following matrix $\Lambda$:
\bea
\Lambda^{ab}=\xi^{\rho}\omega_{\rho}^{ab}+
\n_{[\mu}\xi_{\nu]}e^{a\mu}e^{b\nu},
\la{bt}
\eea
where $\omega_{\rho}^{ab}$ is a spin connection.
For the $AdS$ background we choose $e_0^0=\frac{1}{x_0}$, $e_0^i=e_i^0=0$ and
$e_i^j=\frac{1}{x_0}\d_i^j$ and, therefore,
$\omega_{\mu}^{ij}=0$, $\omega_{\mu}^{0i}=\frac{1}{x_0}\d_\mu^i$.
Computing (\ref{bt}), one finds $\Lambda^{0i}=\p_0\xi^i$ and
$\Lambda^{ij}=\eta^{k[i}\p_k\xi^{j]}$.

A spinor $\psi$ on the $AdS$ background transforms under isometries
just as a scalar. However, if we combine isometries with the
local Lorentz transformations generated by $\Lambda^{ab}$ (\ref{bt}),
the transformation law for $\psi$ modifies as follows
\bea
\nonumber
\d\psi=\xi^{\mu}\p_{\mu}\psi-\frac{1}{4}\Lambda^{ab}\gamma_{ab}\psi.
\eea

With the account of the explicit form of $\Lambda^{ab}$ the last
formula acquires the form
\bea
\la{bcm}
\d\psi=\xi^{i}\p_{i}\psi+\xi^0\p_0\psi-\frac{1}{2}\p_0\xi^i\gamma_{0i}\psi
-\frac{1}{4}\eta^{k[i}\p_k\xi^{j]}\gamma_{ij}\psi.
\eea
Projecting (\ref{bcm}) on $\psi_{\mp}$, one gets
\bea
\nonumber
\d\psi_{\mp}=\xi^{i}\p_{i}\psi_{\mp}+\xi^0\p_0\psi_{\mp}\pm\frac{1}{2}\p_0\xi^i\gamma_i\psi_{\pm}
-\frac{1}{4}\eta^{k[i}\p_k\xi^{j]}\gamma_{ij}\psi_{\mp}.
\eea
As was shown in \cite{HS,MV2} a solution of the Dirac equation vanishing
at infinity has the form
\bea
\la{FS}
\psi_{\pm}(x_0,\vec{k})=\left(\frac{x_0}{\eps}\right)^{\frac{d+1}{2}}
\frac{K_{m\mp \frac{1}{2}}(x_0k)}{K_{m\mp \frac{1}{2}}(\eps k)}\psi_{\pm}(\vec{k}).
\eea
In addition one has the following relation between $\psi_+$ and $\psi_-$:
\bea
\la{rel}
\psi_+(x_0,\vec{k})=-i\frac{\gamma^ik_i}{k}\frac{K_{m-\frac{1}{2}}(kx_0)}
{K_{m+\frac{1}{2}}(kx_0)}\psi_-(x_0,\vec{k}).
\eea
For the sake of clarity we restrict the subsequent consideration
to the case $m>0$. Then in the limit $\eps\to 0$ from (\ref{rel})
one has $\psi_+(\vec{x})=\eps\gamma^i\p_i \psi_-(\vec{x}) +O(\eps)$, while
(\ref{FS}) allows one to determine the boundary value of $\p_0\psi_{\pm}$:
\bea
\p_0\psi_{\pm}(\eps,\vec{x})=
\frac{1}{\eps}\left(\frac{d+1}{2}-m\pm\frac{1}{2}\right)\psi_{\pm}(\vec{x})+O(1),
\eea
Thus, the transformations of $\psi_{\pm}$ induced on boundary acquire the
form
\bea
\nonumber
\d\psi_-(\eps,\vec{x})&=&
\xi^{i}\p_{i}\psi_-+ w(A_kx^k+D)\psi_-
-\frac{\eps^2}{2}\gamma^i\gamma^j A_i\p_j\psi_-
-\frac{1}{4}\eta^{k[i}\p_k\xi^{j]}\gamma_{ij}\psi_-+O(\eps), \\
\nonumber
\d\psi_+(\eps,\vec{x})&=&
\xi^{i}\p_{i}\psi_+ +(w+1)(A_kx^k+D)\psi_+
-\frac{1}{2}\frac{1}{\Box}\gamma^i\gamma^j A_i\p_j\psi_+
-\frac{1}{4}\eta^{k[i}\p_k\xi^{j]}\gamma_{ij}\psi_+ +O(\eps),
\eea
where $w=\frac{d}{2}-m$.
Therefore, on the boundary $\psi_-$ and $\psi_+$ transform in completely
different ways. In particular, in the limit $\eps\to 0$
we recover for $\psi_-$ the standard transformation
rule for fermion with the Weyl weight $w$ under
the conformal mappings:
\bea
\nonumber
\d\psi_-(\vec{x})=
\xi^{i}\p_{i}\psi_-+w(A_k x^k+D)\psi_-
-\frac{1}{4}\eta^{k[i}\p_k\xi^{j]}\gamma_{ij}\psi_-,
\eea
while $\psi_+(\vec{x})$ realizes a nonlocal representation
of the conformal group. Since in CFT quasi-primary operators ${\cal O}(x)$
transform in local representations of the conformal
group only $\psi_-$ can be coupled on the boundary with a primary
operator of the conformal weight $\Delta=\frac{d}{2}+m$.
Thus, the transformation properties of $\psi_{\pm}$ imply that
in the case $m> 0$ we should treat $\psi_-$ and $\bar{\psi}_+$
(analogously for $\bar{\psi}_-$ and $\psi_+$)
as the canonically conjugate coordinate and momentum respectively.

As was discussed in the Introduction in the Hamiltonian version
of the AdS/CFT correspondence one should use an action of the form
$\int(p\dot{q}-H(p,q))$. Action $S_{\psi}$ rewritten in such a form
looks as
\bea
\nonumber
S_{\psi}(\alpha)&=&\int_M d^{d+1}x\sqrt{-G}\left(
-x_0(\bar{\psi}_+\p_0\psi_- + \p_0\bar{\psi}_-\psi_+)
+\frac{\alpha}{2}\bar{\psi}\Gamma^iD_i\psi-\frac{\beta}{2}D_i\bar{\psi}\Gamma^i\psi
-m\bar{\psi}\psi \right. \\
\la{sp1}
&+&\left.\frac{d}{2}(\alpha \bar{\psi}_-\psi_++\beta \bar{\psi}_+\psi_-) \right)
+\frac{1}{2}\int_M d^{d+1}x
\p_0(\sqrt{-G}x_0(\alpha \bar{\psi}_-\psi_++\beta \bar{\psi}_+\psi_-)).
\eea
In passing to the Hamiltonian formulation the total derivative term
in (\ref{sp1}) should be omitted.
Taking into account that
$$
\int_M d^{d+1}x\sqrt{-G}\left(
\frac{\alpha}{2}\bar{\psi}\Gamma^iD_i\psi-\frac{\beta}{2}D_i\bar{\psi}\Gamma^i\psi
\right)=\int_M d^{d+1}x\sqrt{-G}\left(\bar{\psi}\Gamma^iD_i\psi
-\frac{\beta d}{2}(\bar{\psi}_+\psi_- - \bar{\psi}_-\psi_+)\right)
$$
we finally obtain the action that should be used in computing the Green
functions:
\bea
{\bf S}_{\psi}&=&\int_M d^{d+1}x\sqrt{-G}\left(
-x_0(\bar{\psi}_+\p_0\psi_- + \p_0\bar{\psi}_-\psi_+)
+\bar{\psi}\Gamma^iD_i\psi-m\bar{\psi}\psi +d \bar{\psi}_-\psi_+\right).
\la{sp2}
\eea
Note that this action does not depend on $\alpha$ and $\beta$ as one could
expect from the very beginning.

In the Lagrangian picture the total derivative term can be compensated
by adding to action (\ref{sp}) the following boundary term
\bea
\la{fboun1}
I_{\psi}&=&\frac{1}{2}\int_{\p M}
d^dx\sqrt{-\bar{g}}(\alpha \bar{\psi}_-\psi_++\beta \bar{\psi}_+\psi_-).
\eea

For the free theory under consideration the on-shell value of the 
boundary term is equal to 
\bea 
\la{fboun} 
I_{\psi}&=&\frac{1}{2}\int_{\p M} d^dx\sqrt{-\bar{g}}\bar{\psi}\psi.  
\eea
It can be seen as follows. Since ${\bf S}_{\psi}$ does not
depend on $\alpha$ one gets
\bea
\nonumber
{\bf S}_{\psi}&=&S_{\psi}(\alpha)+
\frac{1}{2}\int_{\p M}
d^{d}x \sqrt{-\bar{g}}(\alpha \bar{\psi}_-\psi_++\beta 
\bar{\psi}_+\psi_-) 
=
S_{\psi}(1)+
\frac{1}{2}\int_{\p M}
d^{d}x \sqrt{-\bar{g}}\bar{\psi}\psi. 
\eea 
Now taking into account that on shell $S_{\psi}(\alpha)=0$ 
for any $\alpha$ one obtains the desired equality.
The equality of (\ref{fboun1}) and (\ref{fboun}) seems to hold
in the case of an interacting theory as well.

In \cite{HS} (see also \cite{MV2}) the boundary term (\ref{fboun})
with arbitrary numerical coefficient was suggested as the necessary addition to the Dirac
action (\ref{sp}) to produce the conformally invariant two-point
function of the fermion fields in the boundary CFT.
It comes from our analysis that the Hamiltonian version of
the AdS/CFT correspondence not only reproduces this boundary term but
predicts its numerical coefficient.

It remains to note that when $m<0$ the chiral components $\psi_{\pm}$
change places, i.e. $\psi_+$ becomes a coordinate while $\psi_-$
should be treated as the conjugate momentum. The boundary term
will be given by the same expression (\ref{fboun}) but with the opposite sign.
In the case $m=0$ either $\psi_+$ or $\psi_-$ can be chosen as the
coordinate and the sign of the boundary term depends only on this choice.

Thus, we have shown that using the Hamiltonian version of the AdS/CFT
correspondence we can find all boundary terms necessary in the
Lagrangian formulation. Our consideration can be easily extended to
a supergravity model containing  tensor fields in its spectrum.
In particular, an interesting problem is to find all boundary terms
for type $IIB$ supergravity on the $AdS_5\times S_5$ background.
To this end one can use the covariant action found in \cite{ALS,ALT}.
The solution of this problem would be the first step in establishing
the conformal invariance of higher-point Green functions computed
from the $AdS$ supergravity.

\vskip 1cm

{\bf ACKNOWLEDGMENT} The authors thank I.Y.Aref'eva and L.O.Chekhov
for valuable discussions. We are also grateful to K.Sfetsos
for drawing our attention to the revised version of \cite{MV2} and
for pointing out a misprint in the first version of our paper.
This work has been supported in part
by the RFBI grants N96-01-00608 and N96-01-00551.

\newpage

\end{document}